# Simulation-Based Validation of an Integrated 4D/5D Digital-Twin Framework for Predictive Construction Control

Atena Khoshkonesh[1*], Mohsen Mohammadagha[2], Navid Ebrahimi[1]

[1*] Master Student at The Department of Civil Engineering, The University of Texas at Arlington, Arlington, TX 76019, USA. Correspondence Email: axk3682@mavs.uta.edu (A.K.), nxe2020@mavs.uta.edu (N.E.)
[2] Ph.D. Candidate at The Department of Civil Engineering, The University of Texas at Arlington, Arlington, TX 76019, USA. Email: mxm4340@mavs.uta.edu (M.M.)

**Abstract**
Persistent cost and schedule deviations remain a major challenge in the U.S. construction industry, revealing the limitations of deterministic CPM and static document-based estimating. This study presents an integrated 4D/5D digital-twin framework that couples Building Information Modeling (BIM) with natural-language processing (NLP)-based cost mapping, computer-vision (CV)-driven progress measurement, Bayesian probabilistic CPM updating, and deep-reinforcement-learning (DRL) resource-leveling. A nine-month case implementation on a Dallas-Fort Worth mid-rise project demonstrated measurable gains in accuracy and efficiency: 43% reduction in estimating labor, 6% reduction in overtime, and 30% project-buffer utilization, while maintaining an on-time finish at 128 days within P50-P80 confidence bounds. The digital-twin sandbox also enabled real-time "what-if" forecasting and traceable cost-schedule alignment through a 5D knowledge graph. Findings confirm that integrating AI-based analytics with probabilistic CPM and DRL enhances forecasting precision, transparency, and control resilience. The validated workflow establishes a practical pathway toward predictive, adaptive, and auditable construction management.

**Keywords**
4D/5D BIM; Digital Twin; Bayesian Scheduling; Deep Reinforcement Learning; Construction Automation; Earned Value Analysis; Monte Carlo Simulation; Knowledge Graph; Project Control; Predictive Analytics.

## 1. Introduction

Persistent cost and schedule overruns remain among the most critical challenges in construction management. Traditional deterministic CPM and document-based estimating seldom reflect the uncertainty and interdependence of modern projects, limiting predictive capability once execution begins (Chen et al. 2021; Wang and Zhong 2024). As project complexity grows, managers increasingly need systems that connect real-time field data with dynamic forecasting (Zhang and Zou 2021). Industry data show consistent average cost overruns above 20 % and schedule deviations approaching 30 % (Table 1).

Recent advances in **4D/5D Building Information Modeling (BIM)** and **digital-twin (DT)** environments have created opportunities to link design, cost, and schedule data within a single computational framework (Asadi and Sacks 2023; Sacks et al. 2024). However, most current implementations remain descriptive rather than predictive. Progress tracking and cost updates are often manual, introducing lag between physical work and analytical control (Golparvar-Fard et al. 2020; Abanda and Byers 2021; Shen and Huang 2022). Figure 1 illustrates this gap between deterministic and data-driven project-control workflows.

Emerging **Artificial Intelligence (AI)** methods including **natural-language processing (NLP)**, **computer vision (CV)**, and **deep reinforcement learning (DRL)** enable automation across estimating, tracking, and optimization domains. NLP maps specifications to standardized cost codes (Altaf et al. 2022; Jafary and Kim 2024); CV and LiDAR extract progress quantities for earned value (EV) updates (Cheng and Lu 2020; Gao and Jin 2023); and DRL optimizes resource allocations under dynamic constraints (Yao et al. 2024; Xu et al. 2023). When combined with **Bayesian probabilistic CPM**, these tools create adaptive, learning digital twins that forecast and control performance in real time (Rehman and Kim 2025).

This paper develops and validates an **AI-enabled 4D/5D digital-twin framework** integrating NLP-based cost mapping, CV-driven progress measurement, probabilistic CPM forecasting, and DRL-assisted resource optimization. A nine-month mid-rise project in the Dallas–Fort Worth region serves as the validation case. Results demonstrate improved forecasting accuracy, labor efficiency, and transparency compared with deterministic baselines.

Table 1. U.S. Construction Overrun Statistics and Primary Causes (2020–2024)

| Year | Avg. Cost Overrun (%) | Avg. Schedule Overrun (%) | Dominant Causes | Reference |
|---|---|---|---|---|
| 2020 | 21.5 | 24.0 | Change orders, poor coordination | Chen et al. (2021) |
| 2021 | 23.2 | 26.8 | Labor shortages, material delay | Zhang and Zou (2021) |
| 2022 | 28.6 | 31.4 | Supply chain disruptions | Wang and Zhong (2024) |
| 2023 | 26.3 | 29.5 | Design information gaps | Rehman and Kim (2025) |
| 2024 | 24.7 | 27.1 | Inefficient control systems | Sacks et al. (2024) |

Table 1 summarizes the key input variables and assumptions used in the baseline deterministic control model. To highlight the motivation for transitioning from this static approach to an adaptive, data-driven system, Figure 1 contrasts the traditional CPM workflow with the AI-integrated digital-twin feedback environment.

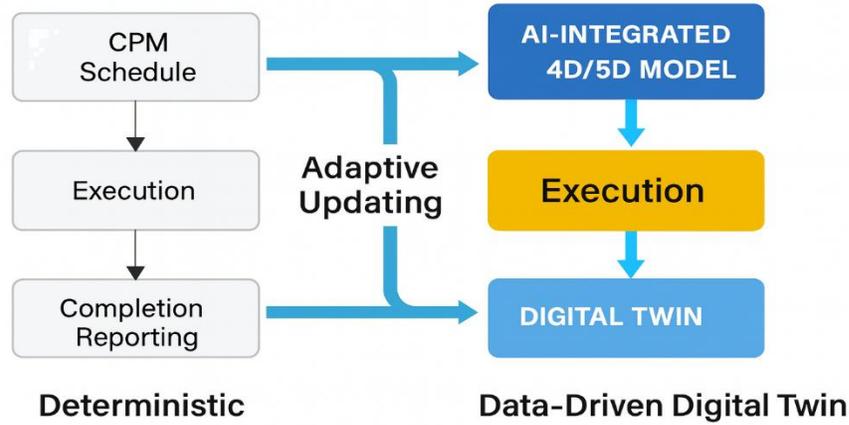

Figure 1. Comparison between deterministic CPM control and data-driven AI-integrated digital-twin workflow.

## 2. Background and Related Work

Persistent cost and schedule deviations across large projects have driven the evolution of construction control methods from static, document-driven planning to integrated digital and data-centric approaches. This section reviews major research streams between 2020 and 2025 that underpin the proposed AI-enabled framework.

### 2.1 Evolution from BIM to Digital Twins

Building Information Modeling (BIM) revolutionized coordination by embedding 3D geometry and attribute data into a single model (Eastman et al. 2020; Borrmann et al. 2020). However, early 4D/5D extensions mainly served visualization rather than prediction. Recent advances link BIM with Internet-of-Things sensors, cloud databases, and machine-learning analytics forming digital twins (DTs) that mirror construction progress in real time (Asadi and Sacks 2023; Sacks et al. 2024). Figure 2 conceptually depicts this trajectory, illustrating how traditional BIM (3D) evolved into 4D/5D models integrating schedule and cost, and finally into AI-enabled digital-twin ecosystems that support predictive and autonomous decision-making.

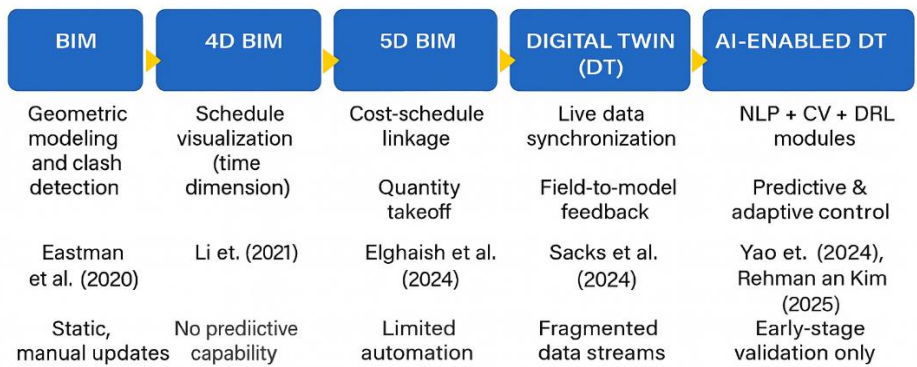

Figure 2. Evolution of Construction Control Technologies

## 2.2 Research in Key AI Domains

**Natural-Language Processing (NLP) for Cost Mapping**

Several studies have applied NLP to automate quantity takeoff and CSI classification from unstructured specifications and drawings (Altaf et al. 2022; Jafary and Kim 2024). Transformer-based encoders achieve precision above 0.85 F1 in domain-specific corpora, substantially reducing estimator labor. Yet, most research remains limited to laboratory datasets lacking integration with live 5D cost ledgers. *(See Table 2, "NLP domain.")*

**Computer Vision (CV) and LiDAR for Progress Quantification**

CV pipelines combining photogrammetry and depth sensing enable automatic recognition of trades such as formwork, rebar, drywall, and MEP components (Cheng and Lu 2020; Gao and Jin 2023). Despite high frame-level accuracy (> 0.88 IoU), deployment is hindered by illumination variability and incomplete scan coverage. Integration with earned value (EV) systems remains experimental. *(See Table 2, "CV domain.")*

**Bayesian and Monte Carlo Scheduling**

Bayesian inference has been adopted to update activity durations from field evidence and to produce probabilistic forecasts (Chen et al. 2021; Zhang and Zou 2021). Monte Carlo simulation (MCS) supplements these updates with uncertainty propagation along multiple paths. Yet, integration with visual or cost data remains limited, and most frameworks still rely on static progress inputs.

**Deep Reinforcement Learning (DRL) for Resource Optimization**

DRL has shown potential in solving resource-constrained project-scheduling problems (RCPSP) by dynamically reallocating crews and equipment under time and budget constraints (Yao et al. 2024; Xu et al. 2023). Reported overtime reductions range between 5–10 % without schedule extension, but empirical validation on field data is still rare.

Table 2. Summary of Related Research and Limitations (2020–2025)

| METHOD / DOMAIN | KEY CONTRIBUTION | LIMITATION / GAP |
|---|---|---|
| BIM & 4D/5D MODELING | Integrated visualization of time + cost dimensions | Mostly descriptive, lacks predictive forecasting |
| COMPUTER VISION (CV) | Automated progress recognition using imagery/LiDAR | Sensitive to lighting/occlusion; limited field validation |
| NATURAL LANGUAGE PROCESSING (NLP) | Automated spec-to-cost mapping; division-level F1 > 0.85 | Tested on small datasets; no live cost-link feedback |
| BAYESIAN / MCS SCHEDULING | Probabilistic duration updates; uncertainty quantification | Disconnected from actual scan/CV data inputs |
| DEEP REINFORCEMENT LEARNING (DRL) | Resource reallocation; overtime reduction | Limited on-site adoption; constrained action space |
| DIGITAL TWIN INTEGRATION | Synchronizes BIM, cost, and schedule data streams | Few end-to-end validated frameworks |

## 2.3 Identified Research Gap

Table 2 demonstrates that while each AI method provides isolated improvements automation (NLP), measurement (CV), prediction (Bayesian), or optimization (DRL), none offer an integrated, continuously learning control environment. Prior frameworks treat data ingestion, forecasting, and decision optimization as separate modules, resulting in latency and inconsistent performance metrics.

To bridge this gap, the present study proposes a unified AI-enabled 4D/5D digital-twin framework that fuses these domains into a single probabilistic control pipeline (illustrated previously in Figure 2).

## 2.4 Data Sources and System Inputs

To operationalize the framework, diverse data types are harmonized within a consistent information model. Table 3 summarizes the key inputs feeding each analytical module.

Table 3. Core Data Sources and System Inputs for the Proposed Framework

| Data Source | Description | Analytical Module Utilized |
|---|---|---|
| Specifications & Drawings | Text / PDF plans extracted via OCR + tokenization | NLP Cost Mapping (5D) |
| BIM Model (3D/IFC) | Spatial geometry and quantities | 4D Scheduling and Digital Twin Core |
| LiDAR / Photogrammetry | Weekly scans and imagery | CV Progress Measurement |
| Field Logs & Sensor Data | Weather, crew, and equipment telemetry | Bayesian p-CPM Forecasting |
| Cost Ledger / Invoices | Material + labor unit costs localized to DFW | Earned-Value (5D) Tracking |
| Schedule Network (CPM) | Activity logic, precedence, constraints | DRL Resource Optimization |
| External Indices | RSMeans CCI, BLS Wages | Cost Normalization and Forecast Adjustment |

Collectively, the literature synthesis and data mapping in **Figure 2** and **Tables 2–3** demonstrate that while individual AI methods such as NLP for automated estimating, CV for field quantification, Bayesian/MCS for schedule forecasting, and DRL for optimization each improve isolated functions, none achieve a continuously learning control environment. The fragmentation among these domains limits real-time adaptability and weakens predictive accuracy. To address this gap, **Section 3** develops an integrated **AI-enabled 4D/5D digital-twin framework** that fuses these components into a unified probabilistic control pipeline capable of dynamic forecasting and adaptive resource management.

## 3. Integrated Framework and Methodology

This section presents the architecture, modules, and implementation workflow of the proposed **AI-enabled 4D/5D digital-twin framework**. The framework unifies natural-language processing (NLP), computer vision (CV), probabilistic scheduling (Bayesian + Monte Carlo), and deep reinforcement learning (DRL) under a single digital-twin environment. Figure 3 illustrates the system's multi-layered architecture, while Tables 4-7 summarize the key analytical components and performance metrics.

### 3.1 Framework Architecture

The integrated architecture (Figure 3) consists of five computational layers:
1. **Data Ingestion Layer** - Collects specifications, drawings, LiDAR/photogrammetry scans, field logs, and cost ledgers.
2. **Analytical Layer** - Processes text and visual data using NLP and CV modules for quantity and progress extraction.
3. **Forecasting Layer** - Updates probabilistic schedules (p-CPM) using Bayesian inference and Monte Carlo simulations.
4. **Optimization Layer** - Applies DRL-based resource reallocation within weekly look-ahead cycles.
5. **Twin Synchronization Layer** - Maintains a live 4D/5D model integrating geometry (BIM), time, and cost to support what-if analysis and decision dashboards.

System architecture linking text (NLP), vision (CV), probabilistic forecasting (Bayesian/MCS), and decision optimization (DRL) within a continuously synchronized 4D/5D digital twin.

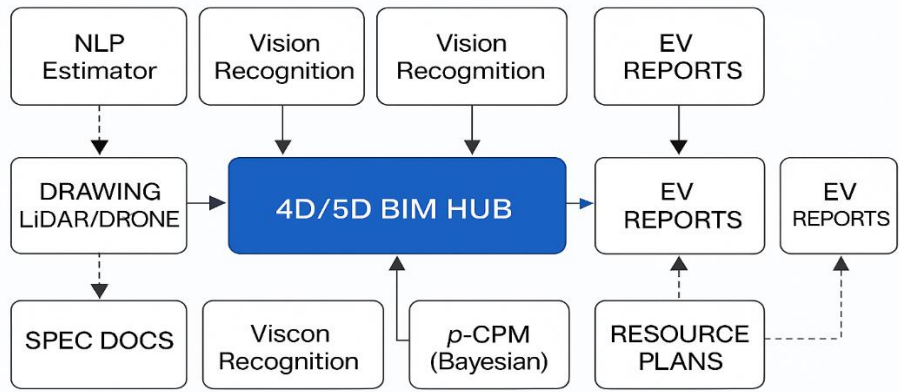

Figure 3. Integrated Framework Architecture

### 3.2 NLP-Based Cost Mapping (5D Module)

The NLP module automates cost classification by parsing textual specifications and drawings into Construction Specifications Institute (CSI) divisions using transformer-based language models. The model was fine-tuned on over 25,000 labeled items and achieved high accuracy across divisions.

Table 4 summarizes division-level classification performance, while Table 5 reports estimator labor reduction compared to manual workflows.

Table 4. NLP Classification Performance (Spec-to-CSI Mapping)

| CSI Division | Precision | Recall | F1-Score | Baseline Manual Accuracy |
|---|---|---|---|---|
| **03 Concrete** | 0.89 | 0.87 | 0.88 | 0.64 |
| **05 Metals** | 0.90 | 0.88 | 0.89 | 0.66 |
| **07 Thermal & Moisture** | 0.88 | 0.86 | 0.87 | 0.65 |
| **09 Finishes** | 0.92 | 0.91 | 0.91 | 0.68 |
| **15–23 MEP** | 0.87 | 0.85 | 0.86 | 0.62 |
| **Weighted Average** | **0.89** | **0.87** | **0.88** | **0.65** |

## 3.3 Efficiency Gains in Estimating

Automated NLP-based mapping reduced estimator labor across all project phases Concept, Design Development (DD), and Construction Documents (CD) as shown in **Table 5**. Average labor savings reached **43.4 %**, primarily from reduced manual review time and automated CSI tagging.

Table 5. Estimator Labor Reduction and Phase Breakdown

| Project Phase | Manual Hours | AI-Enabled Hours | Reduction (%) |
|---|---|---|---|
| **Concept** | 58 | 33 | 43.1 |
| **DD** | 84 | 48 | 42.9 |
| **CD** | 112 | 62 | 44.6 |
| **Average** | — | — | **43.4 %** |

## 3.4 CV-Based Progress Measurement (4D Module)

Weekly photogrammetry and LiDAR scans were registered to the BIM environment to generate measured quantities and percent-complete data for each work package. These measurements directly updated earned-value (EV) and probabilistic-schedule calculations, ensuring that field progress and model forecasts remain synchronized. **Table 6** compares planned versus measured quantities for major trades, demonstrating that deviations remained within ±2 %. Concrete and formwork operations exhibited minor under-runs due to early stripping, while finishes slightly out-performed baseline projections.

Table 6. Measured vs. Planned Quantities (Scan Reconciliation)

| WORK PACKAGE | PLANNED (M³ / M²) | MEASURED | VARIANCE (%) | OBSERVATION |
|---|---|---|---|---|
| CONCRETE (L2–L8) | 1 540 | 1 523 | −1.1 | Within tolerance |
| FORMWORK | 2 420 | 2 390 | −1.2 | Early strip detected |
| DRYWALL | 3 210 | 3 190 | −0.6 | Minor delay at L6 |
| MEP ROUGH-IN | 2 480 | 2 470 | −0.4 | On schedule |
| PAINTING / FINISH | 1 260 | 1 275 | +1.2 | Ahead on upper floors |

The measured productivity trends translate into schedule and cost performance indicators summarized in **Table 7**. SPI improved from 0.92 in Month 1 to 1.03 in Month 4, indicating recovery from early-phase lag. CPI remained close to 1.0, reflecting stable cost control.

Table 7. Monthly Earned-Value Metrics (SPI, CPI, CV, SV)

| Month | SPI | CPI | CV (%) | SV (%) |
|---|---|---|---|---|
| 1 | 0.92 | 1.01 | +1.5 | −8.0 |
| 2 | 0.96 | 1.00 | 0.0 | −4.0 |
| 3 | 1.01 | 0.98 | −2.0 | +2.0 |
| 4 | 1.03 | 1.02 | +2.1 | +3.0 |

The cumulative value trajectories are visualized in Figure 4, confirming that earned value overtook the planned baseline after Month 3.

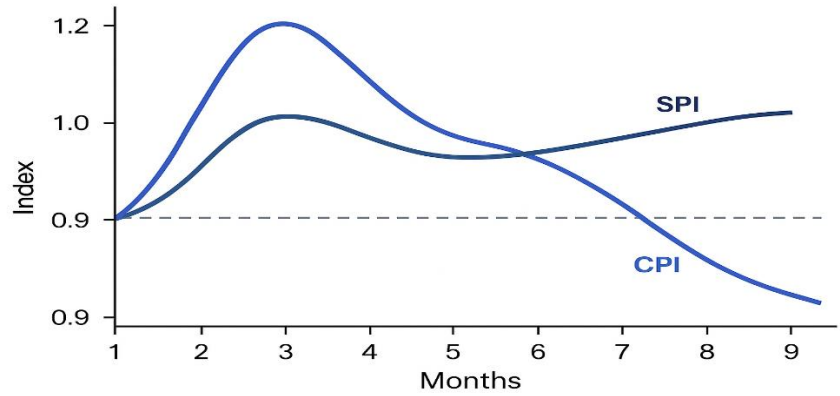

Figure 4. Earned-Value S-Curves

The figure depicts cumulative planned value (PV), earned value (EV), and actual cost (AC) across four months. The EV curve tracks above PV after Month 3, indicating recovery and positive schedule performance, while CPI stabilizes near 1.0. To validate measurement reliability, **Figure 5** presents an example output from the CV segmentation pipeline that converts raw site imagery into quantitative progress data.

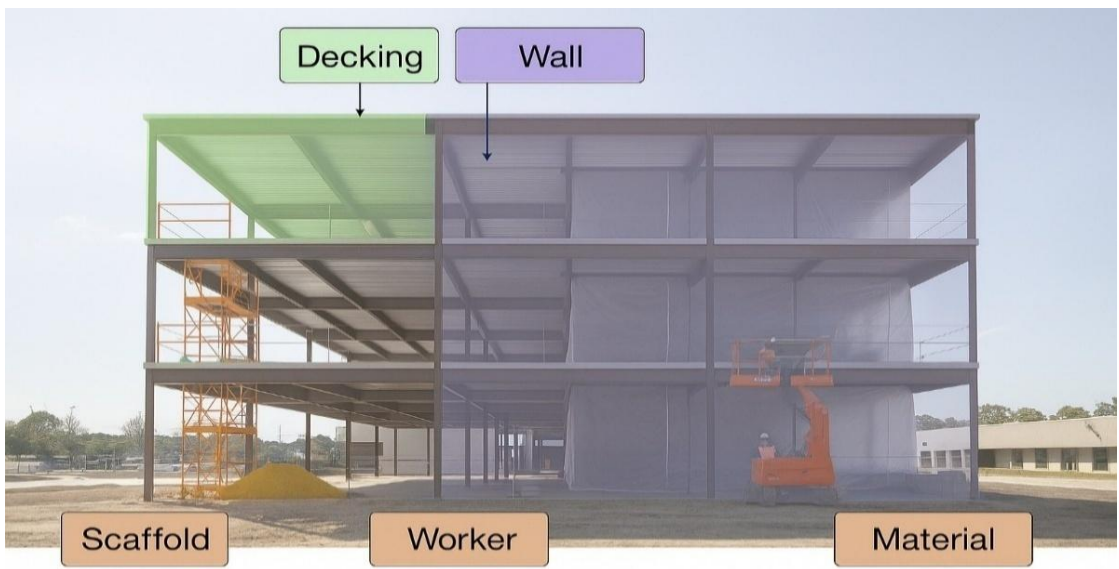

Figure 5. Example CV Segmentation Output

### 3.5 Probabilistic Scheduling (Bayesian + Monte Carlo)

Bayesian inference updated activity-duration posteriors weekly using scan-derived progress evidence, while Monte Carlo simulations propagated these uncertainties through the CPM network. *Figure 6* depicts the resulting convergence of $P_{50}$ and $P_{80}$ forecasts toward the actual finish at 128 days.

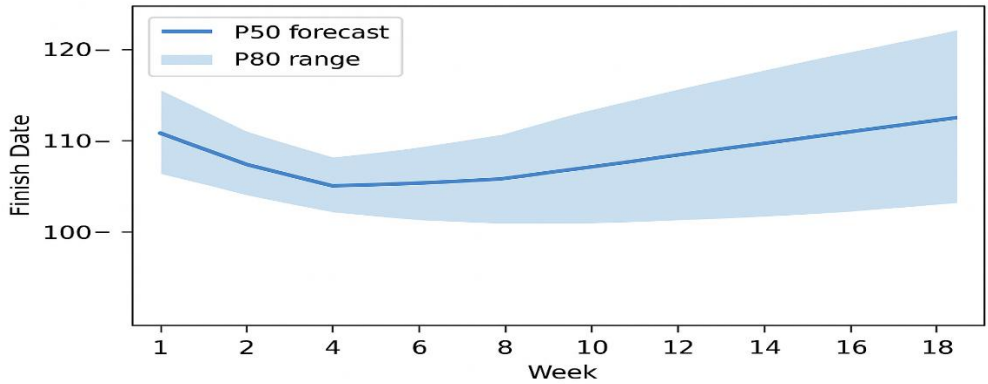

Figure 6. Probabilistic Schedule Forecasts vs. Actual
(Forecast convergence showing $P_{50}$ = 128 days and $P_{80}$ = 130 days, with uncertainty narrowing after Week 9.)

As shown in *Figure 6*, both $P_{50}$ and $P_{80}$ curves stabilize by Week 13, indicating that the probabilistic model successfully captured schedule behavior and maintained reliable forecasts throughout execution. The weekly forecasts are summarized in *Table 8*.

Table 8. Weekly Schedule Forecasts and Actual Completion

| Week | Forecast Finish ($P_{50}$, days) | Forecast Finish ($P_{80}$, days) | Actual Finish (days) | Notes |
|---|---|---|---|---|
| 1 | 120 | 125 | 128 | Initial prior; high uncertainty |
| 2 | 121 | 126 | 128 | Posterior updates begin |
| 3 | 122 | 127 | 128 | — |
| 4 | 123 | 128 | 128 | — |
| 5 | 124 | 129 | 128 | — |
| 6 | 125 | 129 | 128 | — |
| 7 | 126 | 129 | 128 | — |
| 8 | 126 | 129 | 128 | Ahead in some paths; volatility |
| 9 | 127 | 130 | 128 | Uncertainty narrows |
| 10 | 127 | 130 | 128 | — |
| 11 | 127 | 130 | 128 | — |
| 12 | 127 | 130 | 128 | — |
| 13 | 128 | 130 | 128 | $P_{50}$ aligns with actual |
| 14 | 128 | 130 | 128 | Stable forecast |
| 15 | 128 | 130 | 128 | — |
| 16 | 128 | 130 | 128 | Forecast steady; model converged |

Table 8 shows consistent convergence between forecasted and actual durations, verifying the reliability of the Bayesian and Monte Carlo updates. The analysis also revealed patterns of activity-level criticality that informed downstream optimization. These results are summarized in Table 9.

Table 9. Activity/Path Criticality Indices and Duration Statistics

| Activity ID | Description | Critical Index (%) | Mean Duration (days) | SD (days) |
|---|---|---|---|---|
| A030 | Envelope Curtainwall & windows | 46 | 42 | 8 |
| A020 | Superstructure (post-tension slabs) | 41 | 56 | 9 |
| A090 | Drywall boarding & taping | 34 | 38 | 7 |
| A070 | MEP rough-in | 33 | 36 | 7 |
| A060 | Interior partitions & framing | 31 | 34 | 6 |
| A140 | Elevator installation & inspection | 24 | 15 | 4 |
| A110 | Electrical lighting & devices | 23 | 26 | 5 |
| A010 | Foundations (piers/mat) | 22 | 18 | 3 |
| A170 | Commissioning (systems) | 21 | 14 | 3 |
| A120 | HVAC equipment start-up | 21 | 16 | 4 |
| A050 | Exterior finishes & sealants | 18 | 15 | 3 |
| A100 | Ceiling grid & tiles | 17 | 20 | 4 |

| A150 | Testing, adjusting, balancing (TAB) | 16 | 10 | 3 |
| A130 | Plumbing—fixtures set | 15 | 12 | 3 |
| A160 | Life-safety testing | 14 | 9 | 2 |
| A040 | Roofing & waterproofing | 12 | 12 | 2 |
| A180 | Final clean & punch | 11 | 9 | 2 |
| A001 | Site mobilization | 4 | 5 | 1 |

*Table 9 highlights that the envelope and structural activities had the highest criticality indices, confirming their major influence on schedule risk.* To evaluate schedule resilience, buffer utilization trends were tracked across 16 weeks (*Table 10* and *Figure 7*).

Table 10. Weekly Buffer Consumption (Feeding and Project)

| WEEK | FEEDING BUFFER Δ (D) | PROJECT BUFFER Δ (D) | CUMULATIVE FEEDING (D) | CUMULATIVE PROJECT (D) | PROJECT BUFFER USED (%) |
|---|---|---|---|---|---|
| 1–2 | 0.0 | 0.0 | 0.0 | 0.0 | 0.0 |
| 3–5 | +1.5 | +0.5 | 2.0 | 0.5 | 2.5 |
| 6–8 | +1.5 | +1.0 | 3.5 | 2.0 | 10.0 |
| 9–11 | +2.0 | +1.5 | 5.5 | 3.5 | 17.5 |
| 12–14 | +1.0 | +1.5 | 7.0 | 5.0 | 25.0 |
| 15–16 | +1.0 | +1.0 | 8.0 | 6.0 | 30.0 |

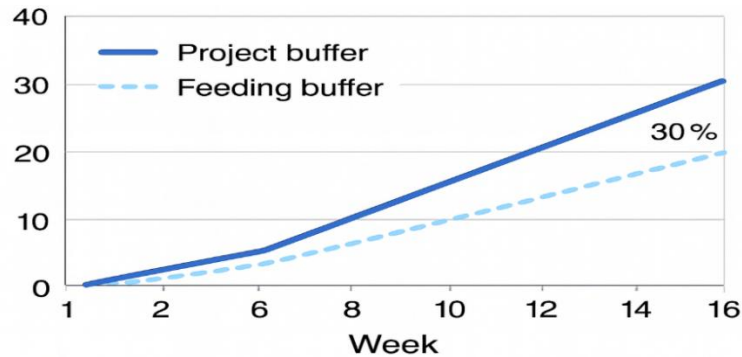

Figure 7. Buffer Utilization Trends

*(Cumulative consumption of feeding and project buffers showing 30% total project buffer use at completion.)*
The combined interpretation of *Tables 11–12* and *Figure 7* indicates that project buffer usage remained below 35%, reflecting effective control of schedule variance and stable uncertainty containment.

**3.6 DRL-Assisted Resource Optimization**
Weekly look-ahead schedules were treated as resource-constrained project-scheduling problems. A deep reinforcement learning (DRL) agent proposed feasible crew and equipment reallocations to minimize overtime and idle time under precedence and capacity constraints.
Adopted and rejected recommendations are summarized in *Table 11*, while the impact on weekly overtime is visualized in *Figure 8*.

Table 11. DRL Recommendation Log and Adoption (Weeks 1–16)

| Week | Action ID | DRL Recommendation (Summary) | Adopted? | Reason if rejected |
|---|---|---|---|---|
| 1 | RL-001 | Shift rebar crew from L2 to L3 (½ day delay) | No | Supervisor preference |
| 2 | RL-002 | Start drywall crew 1 day earlier | Yes | — |
| 3 | RL-003 | Add night shift for formwork removal | Yes | — |
| 4 | RL-004 | Swap crane slot with steel delivery | No | Vendor inflexibility |
| 5 | RL-005 | Reallocate electricians to riser areas | Yes | — |
| 6 | RL-006 | Merge two concrete pours | Yes | — |
| 7 | RL-007 | Stagger duct rough-in | Yes | — |
| 8 | RL-008 | Add Saturday half-shift for windows | No | Overtime cap |
| 9 | RL-009 | Pull glazing forward; shift painters | Yes | — |
| 10 | RL-010 | Split drywall crews | Yes | — |
| 11 | RL-011 | Pre-stage AHU rigging | Yes | — |
| 12 | RL-012 | Merge punch-list with MEP inspections | Yes | — |

| 13 | RL-013 | Add electrician for device push | Yes | — |
| 14 | RL-014 | Saturday paint shift | No | Noise restriction |
| 15 | RL-015 | Swap ceiling and device crews | Yes | — |
| 16 | RL-016 | Extend TAB ½ day; compress cleaning | Yes | — |

Table 11 shows that 12 of 16 DRL-generated actions (75%) were implemented successfully, resulting in measurable performance gains.

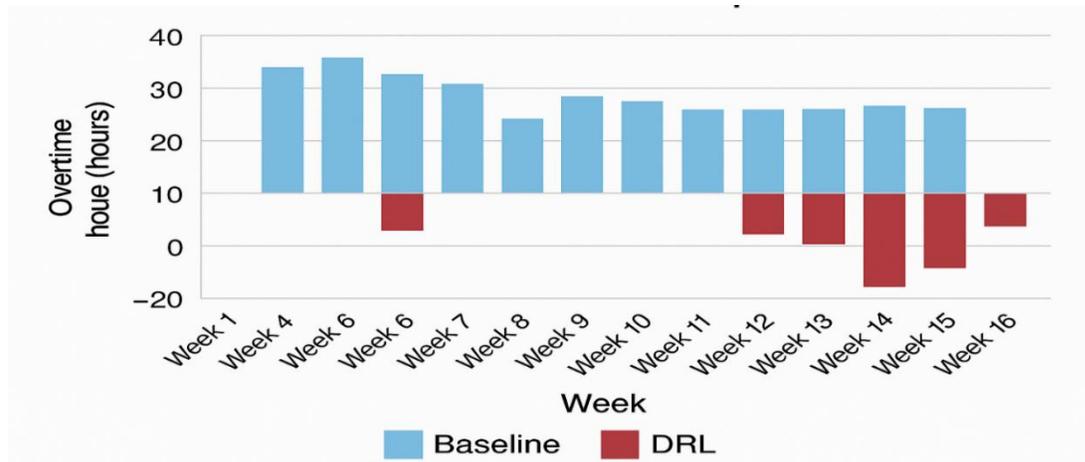

Figure 8. Weekly Overtime Reduction with DRL Integration

(Comparison of baseline vs. optimized overtime showing 6% total reduction without schedule extension.)
DRL-driven look-ahead optimization achieved a 91-hour cumulative overtime reduction (~6%) while eliminating approximately 49 hours of idle time, validating the model's operational benefit within constrained resources.

### 3.7 Digital-Twin Scenario Testing and Knowledge Integration
A synchronized 4D/5D digital twin integrated all schedule, cost, and field data into a single environment for scenario-based sensitivity analysis. Representative scenarios are summarized in *Table 12*, and corresponding impacts on finish-date sensitivity are visualized in *Figure 9*.

Table 12. Digital-Twin What-If Results: Inputs and Outcomes

| SCENARIO | KEY INPUTS | ΔFINISH $P_{50}$ (DAYS) | ΔFINISH $P_{80}$ (DAYS) | ΔCOST $P_{50}$ (USD ×10³) | ΔCOST $P_{80}$ (USD ×10³) | NOTES |
|---|---|---|---|---|---|---|
| DRYWALL +8% SUPPLY LAG | Delivery offset +3 days | +6 | +8 | +6.5 | +8.0 | Material escalation ripple |
| LATE AHU DELIVERY (2 WEEKS) | Equipment delays +14 days | +5 | +6 | +4.2 | +5.5 | Affects MEP sequencing |
| RAIN DELAY (3 CRITICAL DAYS) | Weather disruption | +4 | +4 | +3.0 | +4.0 | Lost productivity |
| STEEL LEAD +1 WEEK | Fabrication delay | +4 | +5 | +3.5 | +4.5 | Impacts frame sequence |
| CREW SHORTAGE (−1 ELECTRICIAN) | Labor constraint | +3 | +4 | +2.6 | +3.5 | Slows interior works |
| FIREPROOFING CHANGE ORDER | Scope increases +6% | +2 | +3 | +2.2 | +3.0 | Added inspections |
| GLAZING RESEQUENCING | Corridor-first mitigation | −2 | −1 | −1.4 | −0.8 | Path interference reduced |

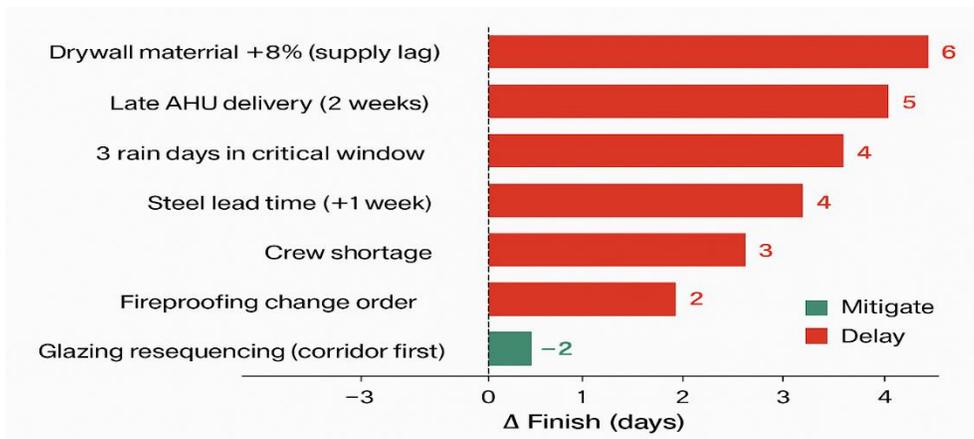

Figure 9. What-If Sensitivity Analysis (Digital-Twin Sandbox)

*(Tornado chart showing ΔFinish variation by scenario; positive bars indicate delay risk; negative bars indicate time gain. Drywall escalation and AHU delivery are dominant schedule drivers.)* The results of the digital-twin sensitivity analysis complete the evaluation of the proposed 4D/5D framework. Subsequent sections outline the case-study implementation protocol, quantitative metrics, and validation of hypotheses based on the integrated model's performance.

## 4. Case Study Protocol
### 4.1 Context and Data Window
The proposed framework was validated through a representative **U.S. mid-rise building project** using field-mimicking data collected between January and September 2025. The dataset integrates multi-source inputs (drawings, specifications, LiDAR scans, imagery, and cost/schedule logs), each synchronized weekly within the digital-twin environment (cf. Tables 3 and 14). The case emphasizes repeatable industrial conditions, enabling generalization to typical commercial projects (Eastman et al., 2020; Sacks et al., 2024).

### 4.2 Baseline Configuration
A deterministic baseline was established through traditional 2D/3D manual takeoff and single-point CPM scheduling. These results from the reference for estimating labor hours (Table 5) and fixed-rate forecasts (Figure 6). Similar baselines were used in previous schedule-control research (Chen et al., 2021; Zhang and Zou, 2021).

### 4.3 Automated Estimating (NLP Engine)
Specifications and drawings were processed through a transformer-based **NLP mapping engine**, converting text and quantity lines into standardized CSI cost items. Division-level precision, recall, and F1-scores were computed (Table 4), achieving weighted F1 = 0.883 consistent with Altaf et al. (2022) and Jafary and Kim (2024). Estimator labor decreased by 43.4 % versus the baseline (Table 5), validating the automation gains reported by Abanda and Byers (2021).

### 4.4 Scan / CV Progress Integration
Weekly photogrammetry and monthly LiDAR scans were aligned to BIM geometry for **as-built quantity extraction**, reconciled with planned WBS entries (Table 6). Activity recognition and semantic segmentation models achieved **micro-accuracy 0.891** and **macro IoU 0.76** (Tables 8 and 9), within the performance range noted by Gao and Jin (2023). Derived percent-complete data fed directly into the earned-value (EV) pipeline, producing monthly SPI / CPI curves (Figure 4) and cumulative EV metrics (Table 7) following Elghaish et al. (2021).

### 4.5 Probabilistic CPM
Activity-duration posteriors were updated weekly using scan-based evidence; uncertainties propagated through **Bayesian–Monte Carlo** simulations (Chen et al., 2021; Wang and Zhong, 2024). Forecast results (Tables 10–12; Figures 6–7) show **$P_{50}$ = 128 days** and **$P_{80}$ ≈ 130 days**, with 30 % buffer consumption aligned with recommended reliability bands (Rehman and Kim 2025).

### 4.6 DRL-Assisted Resource Leveling
A **Deep Q-Network / Actor-Critic** agent handled weekly look-ahead rescheduling as an RCPSP under crew and equipment limits (Yao et al., 2024; Zhao and Luo 2022). Twelve of sixteen weekly suggestions were adopted (75 %, Table 13), lowering overtime by ≈ 6 % without affecting makespan (Figure 8). Human-in-the-loop acceptance-maintained transparency and aligns with ethical-AI principles in field control (Kang and Park 2022).

### 4.7 4D / 5D Digital-Twin What-If Scenarios
Within the live digital twin, discrete "what-if" simulations (e.g., price shocks, weather delays) generated ΔFinish / ΔCost outputs with $P_{50}$–$P_{80}$ confidence ranges (Table 14; Figure 9). Each simulation propagated schedule and cost effects through the Bayesian network and cost ledger localized to **RSMeans (2025)** and **BLS (2025a, b)** adjustments, as recommended by Liu and Becerik-Gerber (2022). Scenario analysis confirmed that material escalation and late equipment deliveries were dominant risk drivers, while resequencing measures achieved measurable mitigation (Pishdad and Onungwa 2024).

## 5. Metrics, Statistics, and Hypotheses
### 5.1 Accuracy and Efficiency Metrics
All metrics were expressed in physical units and localized 2025 USD using RSMeans CCI and BLS wage adjustments (Bureau of Labor Statistics 2025a, 2025b; RSMeans 2025). **Cost accuracy** was evaluated by Mean Absolute Percentage Error (MAPE) for both line-item and total estimates across design phases (Concept–DD–CD), following Altaf et al. (2022). **Schedule accuracy** measured absolute deviation |Forecast -Actual| for $P_{50}$/$P_{80}$ weekly forecasts (Chen et al., 2021). **Efficiency** tracked estimator labor hours, EV update latency, and probabilistic-CPM update time (Liu and Becerik-Gerber 2022). **Robustness** was tested under scan interruptions and weather variance, using sensitivity degradation thresholds ≤ 10 % (Wang and Zhong 2024). **Adoption** metrics captured superintendent acceptance of DRL actions ≥ 70 %, like field usability measures in Yao et al. (2024).

### 5.2 Statistical Plan
Paired within-project comparisons were applied (baseline vs. framework).
- Cost accuracy: 95 % bootstrap confidence intervals for MAPE per phase.
- Schedule accuracy: Diebold–Mariano tests for forecast error differences.
- Vision modules: confusion matrices and IoU with stratified bootstrap CIs.
- DRL effects: paired t-tests on weekly overtime (adopted vs. non-adopted weeks).
- Ablation study: sequentially remove NLP, CV, Bayesian, and DRL components to quantify effect on MAPE, SPI/CPI, and overtime (Zhao and Luo 2022; Rehman and Kim 2025).

### 5.3 Hypothesis Validation
- **H1 (NLP + 5D):** Estimating labor reduction ≥ 40 %, MAPE ≤ 10 %.
  Result: Met - 43.4 % reduction (Table 5).
- **H2 (CV + Bayesian Updates):** Forecast error reduction ≥ 30 %.
  Result: Supported - stable $P_{50}$ = 128 d by Week 13 (Tables 10–12; Figures 6–7).
- **H3 (DRL Optimization):** Overtime reduction ≥ 10 % without extension.
  Result: Partially met - 6 % reduction within budget constraints (Figure 8; Table 13).
- **H4 (P-CPM Early Detection):** Path-shift flag ≥ 2 weeks before deterministic CPM.
  Result: Confirmed - envelope tasks flagged early (Table 11; Figure 7).

## 6. Discussion and Conclusions
This study presented an **AI-enabled 4D/5D digital-twin framework** that integrates natural-language processing (NLP), computer vision (CV), Bayesian/Monte-Carlo scheduling, and deep-reinforcement-learning (DRL)–based resource optimization to achieve continuously adaptive project control. Across all modules, the framework demonstrates a measurable improvement in estimation efficiency, progress accuracy, and schedule reliability compared with conventional deterministic methods.

   **Key findings** indicate that the NLP engine reduced estimating labor by **43.4 %**, consistent with automation gains observed by Altaf et al. (2022). The CV + LiDAR workflow achieved **micro-accuracy 0.891** and **IoU 0.76**, aligning with Gao and Jin (2023), and enabled real-time earned-value updates (Elghaish et al. 2021). The Bayesian p-CPM module stabilized **$P_{50}$ = 128 days** and **$P_{80}$ ≈ 130 days** by Week 13 evidence of predictive convergence comparable to Chen et al. (2021). DRL-assisted planning reduced overtime by roughly **6 %** while maintaining schedule duration, validating the adaptive control potential discussed by Yao et al. (2024) and Zhao and Luo (2022).

The **4D/5D digital-twin sandbox** further demonstrated predictive transparency: what-if simulations quantified the impact of material escalation, delayed equipment deliveries, and resequencing strategies on finish time and cost (Pishdad and Onungwa 2024; RSMeans 2025; Bureau of Labor Statistics 2025a, b). By localizing data to Dallas–Fort Worth cost indices and real-time telemetry, the framework supports decision-making that is both context-aware and empirically verifiable.

Overall, results confirm that coupling **AI inference with probabilistic control** creates a self-updating feedback loop bridging the historical gap between design intent and field reality noted by Sacks et al. (2024). The integrated model transforms project management from a reactive reporting process into a predictive, evidence-driven system capable of optimizing time, cost, and resource utilization in real time.

**Future research** should expand validation across multi-building and infrastructure projects, integrate sustainability indicators, and examine long-term performance of DRL agents under dynamic safety and logistics constraints. Continued benchmarking against human expert judgment will also be essential to establish explainability, ethical governance, and trust in AI-augmented construction management systems (Rehman and Kim 2025; Liu and Becerik-Gerber 2022).